# ADMET polymerization of α,ω-unsaturated glycolipids: synthesis and physico-chemical properties of the resulting polymers†

G. Hibert,[a,b] E. Grau,[a,b] D. Pintori,[c] S. Lecommandoux*[a,b] and H. Cramail*[a,b]

Trehalose diesters exhibiting α,ω-unsaturation are glycolipids which can be easily polymerized by ADMET (acyclic diene metathesis) polymerization. In this paper, enzymatic esterification was performed to selectively esterify primary hydroxyl groups of trehalose (6 and 6'-positions) with vinyl undecenoate. The vinyl ester was beforehand obtained by palladium-catalyzed transesterification of undecenoic acid with vinyl acetate. The resulting trehalose diundecenoate was homopolymerized and copolymerized with undecenyl undecenoate in order to obtain random copolymers with different compositions. The synthesis of such copolymers was confirmed by $^1$H NMR spectroscopy and size exclusion chromatography (SEC). Their solid-state phase separation were investigated by DSC and X-ray scattering as function of temperature and their solution self-assembly was investigated by dynamic light scattering (DLS) in water.

## Introduction

Glycopolymers have been the subject of a large number of studies over the last decades.[1–5] Most of them are obtained from the polymerization of petroleum-based monomers bounded to a sugar moiety, probably excepted recent examples of glycoplypeptides.[6,7] However, other sugar-containing molecules, such as glycolipids, are fully bio-based and could be used for the synthesis of glycopolymers.

Glycolipids are amphiphilic molecules containing a sugar part, which is hydrophilic, linked to a fatty acid moiety, which is hydrophobic.[4,8–11] Among these glycolipids, sugar esters are a large family showing a huge diversity, due to the possible combination of different sugar moieties (mono-, di-, oligo- and polysaccharides) with a wide variety of fatty acid derivatives.[12] In addition, these molecules are biodegradable and exhibit surfactant and bio-active properties.[13–19] Even if glycolipids appear to be good candidates for the synthesis of glycopolymers, only few studies describe the polymerization of sugar esters. Our group has recently reported the synthesis of thermoplastic polymers from methyl α-D-glucoside monoester and sucrose monoester.[20] These sugar esters with hydroxyl groups on the fatty acid chains were polymerized with fatty acid-based diols (4-hydroxybutyl 9-hydroxy-10-ethoxyoctadecanoate) and isophorone diisocyanate. Both linear and cross-linked polyurethanes were obtained depending on the solvent of polymerization.

Park et al. described the synthesis of polyesters from trehalose and sucrose diesters. The sugar ester was prepared by esterification of these carbohydrates with divinyl adipate in acetone in the presence of Novozyme 435 as catalyst.[21] The lipase enabled the selective esterification of the primary hydroxyls of the saccharide. Using an excess of divinyl adipate, the diester could be selectively obtained. Polycondensation reaction was then performed in acetone with different diols to get various degradable glycopolyesters with trehalose and sucrose moieties within the polymer chains.

Gross and coll. have also described the synthesis of novel glycopolyesters from lactonic sophorolipid.[22,23] Lactonic sophorolipid was synthesized from the fermentation of Candida bombicola and polymerized by ring-opening metathesis polymerization (ROMP) using Grubbs catalysts. In another study, the same group reported the ROMP of chemo-enzymatically modified lactonic sophorolipids. Copolymers of these acetylated and modified sophorolipids were obtained using a similar synthetic methodology. Polymers were then post-modified using so-called "click chemistry" reactions on acrylated groups. Functionalization of methacrylate was performed via a thiol–ene reaction with mercaptoethanol and functionalization of the azide was performed via azide–alkyne Huisgen cyclo-addition reaction. The synthesized polymers from sophorolipids were shown to be compatible with human mesenchymal stem cells. Moreover, their biodegradation has been evidenced and can be controlled depending on the functions linked to the sophorose moiety. This study revealed that polymers from sophorolipids could be used as bioresorbable polymers in biomedical applications.[24]

Among the natural glycolipids available, trehalose lipids are a family of glycolipids with trehalose as saccharide moiety. These molecules are 6,6'-trehalose diesters and they are basic components of microorganism (*Mycobacteria* or *Corynebacteria* and *Caenorhabditis elegans* dauer larvae).[25–27] These glycolipids show anti-tumour and antibacterial activities which have been the subject of many studies to understand their biological potential. Several routes have been developed for the synthesis of trehalose esters. To target precise morphologies (trehalose esterified only on the primary hydroxyl groups in 6 and 6' position), routes using protection and deprotection steps of the secondary alcohol of the trehalose were developed. These procedures allow a selective esterification of the primary alcohols but need several reaction steps to get the sugar diesters.[25,28–30] To reduce the number of reaction steps while keeping the selectivity of the esterification reaction, protecting group-free strategies were also developed. Some authors used peptide-coupling agent like TBTU or the couple triphenyl phosphine (PPH$_3$)/ diethyl azodicarboxylate (DIEAD).[31,32] According to the stoichiometry between trehalose and fatty acids, diesters or monoesters of trehalose can be obtained. These routes are however showing some drawbacks, a main one being the solvents used (DMF, pyridine) which are toxic and even classified as CMR substances. Finally, many studies have reported the enzymatic esterification of trehalose. Generally, lipase B from *Candida antarctica* is usually employed. The use of enzymes allows a one step and selective esterification on primary alcohol in mild conditions with low toxic reagent and solvents.[33–35]

In this contribution, the preparation of α,ω-unsaturated trehalose diester from an enzymatic esterification of the trehalose with vinyl undecenoate is reported. The vinyl undecenoate was beforehand synthesized by transvinylation of undecenoic acid with vinyl acetate. Then, this new trehalose diester was homopolymerized and copolymerized with undecenyl undecenoate by acyclic diene metathesis (ADMET) to produce polymers with original features. Finally, the thermal and self-assembly properties of the so-formed amphiphilic random copolymers were investigated respectively in bulk and solution. To our knowledge this is the first time that ADMET was used to polymerize glycolipids.

## Experimental

### Materials

Potassium hydroxide (KOH, pellet), 2-methyltetrahydrofuran (99%), vinyl acetate (99%), Hoveyda-Grubbs 2nd generation metathesis catalyst, 1,5,7-triazabicyclodec-5-ene (TBD, 98%), 2-methyl-tetrahydrofuran (mTHF, 99%), Lipase B from Candida antarctica (CALB) were purchased from Sigma-Aldrich. 10-undecen-1-ol (99%), 10-undecenoic acid were supplied from Alfa Aeser. Methyl 10-undecenoate (99%), palladium acetate (98%), were purchased from TCI. Anhydrous trehalose (99%) was purchased from Fisher.

### Instrumentation

Flash chromatography was performed on a Grace Reveleris apparatus, employing cartridges from Grace equipped with ELSD and UV detectors at 254 and 280 nm. Elution solvents are dependent on the sample and are mentioned in the experimental parts.

All NMR spectra were recorded at 298 K on a Bruker Avance 400 spectrometer operating at 400MHz. Size exclusion chromatography (SEC) measurements were conducted on a PL GPC50 integrated system with RI detectors with a series of three columns from Polymer Laboratories. DMF with LiBr (1 g.L$^{-1}$) was used as eluent at 25°C at an elution rate of 1mL/min. Polystyrene was used as the standard. Differential Scanning Calorimetry (DSC) measurements were performed on DSC Q100 (TA Instruments). The sample was heated from −150°C to 170°C at a rate of 10°C.min$^{-1}$. Consecutive cooling and second heating runs were also performed at 10°C.min$^{-1}$. Thermogravimetric analyses (TGA) were performed on TA Instruments Q50 from room temperature to 600°C at a heating rate of 10°C.min$^{-1}$. The analyses were investigated under nitrogen atmosphere with platinum pans. ESI-MS was performed on a QStar Elite mass spectrometer (Applied Biosystems) and MALDI-TOF spectra were performed on a Voyager mass spectrometer (Applied Biosystems) both by the Centre d'Etude Stucturale et d'Analyse des Molécules Organiques (CESAMO). The ESI-MS instrument is equipped with an ESI source and spectra were recorded in the negative/positive mode. The electrospray needle was maintained at 4500 V and operated at room temperature. Samples were introduced by injection through a 20 μL sample loop into a 400 μL/min flow of methanol from the LC pump. Samples were dissolved in THF at 1 mg/ml, and then 10 μl of this solution was diluted in 1 ml of methanol. The MALDI-TOF instrument is equipped with a pulsed $N_2$ laser (337 nm) and a time-delayed extracted ion source. Spectra were recorded in the positive-ion mode using the reflectron and with an accelerating voltage of 20 kV. Samples were dissolved in DMF at 10 mg/mL. DHB (2,5-dihydroxybenzoic acid) was employed as the matrix for ionization. Transmission Electron Microscopy images were recorded at the Bordeaux Imaging Center (BIC) on a Hitachi H7650 microscope working at 80 kV in high resolution mode. Samples were prepared by spraying a 0.5 mg/mL aqueous solution of polymer nanoparticles onto a copper grid (200 mesh coated with carbon) using a homemade spray tool. No stained were applied in this case. Dynamic light scattering measurements were performed at 25°C with a Malvern Instrument Nano-ZS equipped with a He-Ne laser (l ¼ 632.8 nm). Samples were introduced into cells (pathway: 10 mm) after filtration through 0.45 μm cellulose micro-filters. The measurements were performed at a scattering angle of 90°.

### Synthesis of monomers

**Vinyl undecenoate (1) (transvinylation of undecenoic acid):** Undecenoic acid (1 eq.) and a 15 eq. excess of vinyl acetate (VAc) was poured in a CEM Discover SP microwave reactor vial. Then, the palladium acetate (0.05 eq.), and the potassium hydroxide (0.10 eq.) were added and the resulting reaction mixture was stirred under microwave at 60 °C for 2 h. The reaction mixture was diluted in DCM and then filtrated over celite to remove the palladium acetate, before removing the solvent with a rotary evaporator. The resulting residue was purified by silica gel flash chromatography using an elution gradient of 2-5% MeOH in DCM to give the vinyl undecenoate. Yield: 95 %. $^1$H NMR (DMSO-$d_6$, 400MHz, δ (ppm)): 7.29 (m, 1H, =CH-OCO-), 5.81 (m, 1H, -CH=$CH_2$), 4.97 (m, 2H, $CH_2$=CH-), 4.88 (d, 1H, $CH_2$=CH-OCO), 4.56 (d, 1H, $CH_2$=CH-OCO-), 2.37 (t, 4H, -$CH_2$-COO-), 2.04 (m, 4H, -$CH_2$-CH=CH-), 1.67 (m, 4H, -$CH_2$-$CH_2$-COO-), 1.30 (m, 20H, aliphatic -$CH_2$-).

**Trehalose diundecenoate (2) (enzymatic esterification of trehalose):** The lipase (2.8 g) was added to a mixture of trehalose (3 g, 9 mmol), vinyl ester (6.8 g, 22 mmol, 2.5 eq) in dry mTHF (40 mL). The reaction mixture was stirred at 45 °C for 72 hr. After the reaction time, mTHF was added to well dissolve the diesters of trehalose, then the reaction mixture was filtered and solvent was removed in rotary evaporator. The obtained crude product was purified by silica gel flash chromatography using an elution gradient of 5-25% methanol in EtOAc-DCM (1:1) to give pure trehalose diesters as white solids. Yield: 50 %. $^1$H NMR (DMSO-$d_6$, 400MHz, δ (ppm)): 5.78 (m, 2H, -CH=$CH_2$), 5.04 (d, 2H, -OH, H4), 4.94 (m, 4H, $CH_2$=CH-), 4.89 (d, 2H, -OH, H3), 4.82 (d, 2H, -CH-, H1), 4.76 (d, 2H, -OH,

H2), 4.21 (d, 2H, -CH-, H6), 4.04 (m, 2H, -CH-, H6), 3.89 (m, 2H, -CH-, H5), 3.55 (m, 2H, -CH-, H3), 3.26 (m, 2H, -CH-, H2), 3.13 (m, 2H, -CH-, H4), 2.27 (t, 4H, -CH$_2$-COO-), 2.01 (m, 4H, -CH$_2$-CH=CH-), 1.51 (m, 4H, -CH$_2$-CH$_2$-COO-), 1.33-1.25 (m, 20H, aliphatic -CH$_2$-). (**Fig S1-S2**)

**Undecenyl undecenoate (3) (transesterification):** Undecenol (12.8 g, 0.08 mol.) was blended with 10-methylundecenoate (15 g, 0.08 mol.). TBD (5% mol.) was added as a catalyst. The reaction was performed under a nitrogen flow at 120°C for 2h, then the temperature was increased to 160 °C for 2h more under dynamic vacuum. Purification over silica gel flash chromatography was performed using cyclohexane/ethyl acetate 94/6 eluent. Yield: 76%. $^1$H NMR (400MHz, CDCl$_3$, δ (ppm)): 5.8 (m, 1H, -CH=CH$_2$), 4.9 (m, 2H, CH$_2$=CH-), 4.0 (t, 2H, -CH$_2$COO-), 2.2 (t, 2H, -COOCH$_2$-), 2.0 (m, 4H, -CH$_2$-CH=CH-), 1.5-1.2 (m, 26H, aliphatic -CH$_2$-) (**Fig S3**).

**Synthesis of polymers by ADMET polymerization**

**Homopolymers (P0 and P100):** Into a flame-dried Schlenk flask equipped with bubbler, undecenyl undecenoate (P0) (0.1 g, 0.3 mmol.) or trehalose diundecenoate (P100) (0.2 g, 0.3 mmol.) dried over-night under vacuum was dissolved in 2 mL of dry THF. Hoveyda-Grubbs 2$^{nd}$ generation metathesis catalyst (4 mol.%) was added and the reaction mixture was stirred under nitrogen atmosphere for 24 h at 45 °C. Then, 3 ml of ethyl vinyl ether were introduced into the flask to quench the reaction. The final glycopolyester (P100) or fatty acid-based polyester (P0) was purified by precipitation in cold methanol.

Copolymers (P33, P50 and P67): Into a flame-dried round Schlenk flask, trehalose diundecenoate (0.2 g, 0.3 mmol.) and the corresponding amount of undecenyl undecenoate (see **Table 1**) were mixed and dried over-night under vacuum. Then, the diene monomers were dissolved in 2 mL of dry THF and Hoveyda-Grubbs 2$^{nd}$ generation (4 mol.%) was added. The reaction mixture was stirred at 45 °C under nitrogen atmosphere for 24 h. Then, 3 mL of ethyl vinyl ether and 4 mL of THF were introduced to the flask. The final copolymers were purified by precipitation in cold methanol.

**Polymer self-assemblies in water**

Nanoparticles were prepared using solvent displacement method by dialysis. (Co)polymers were solubilized in DMF and polymer solutions were then introduced into dialysis membranes with molecular weight cutoff of 1 kD. The membranes were beforehand soaked for 15 min and rinsed with deionized water. The membranes were then submerged in 2 L of deionized water and dialyzed under gentle magnetic stirring for 12 h. After 2 h and 4 h, the dialysis solution was replaced by fresh deionized water. At the end of the dialysis, milky solutions present inside the dialysis membranes were recovered.

## Results and discussion

**Synthesis of trehalose diester**

The high selectivity of enzymes and their high activity in mild conditions have raised interest from lot of research groups. In this context, the enzymatic esterification of sugar esters and particularly trehalose esters have been already reported.[34,36–39] Fatty acids or fatty acid methyl (or vinyl) esters can be employed for this esterification.[34] Vinyl esters are especially interested as, acetaldehyde, -a gas at room temperature- that is produced during their transesterification can be readily removed from the reaction medium, thus allowing the displacement of the reaction towards the ester formation. Herein, the enzymatic transesterification of trehalose with vinyl ester has been developed to design new trehalose diesters.

Different catalysts are used in the literature to synthesize vinyl esters from vinyl acetate and fatty acids. The synthesis of vinyl undecenoate (**Scheme 1, (1)**) was carried out by a transvinylation reaction between undecenoic acid and vinyl acetate under Pd(II) catalysis with KOH at 60°C under microwave.

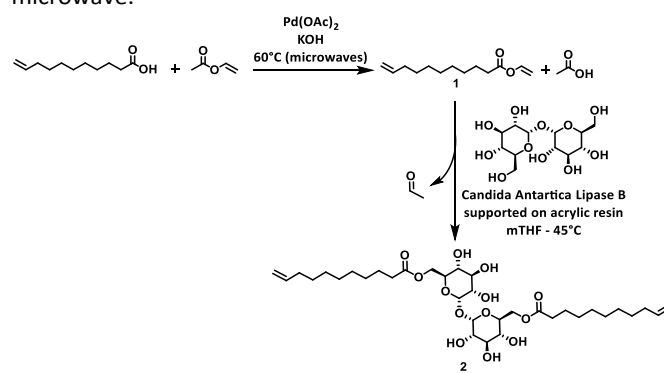

**Scheme 1** Synthesis of vinyl undecenoate (1) and trehalose diundecenoate (2).

The $^1$H NMR spectrum confirms the formation of the vinyl undecenoate (**Fig 1**) with the appearance of the peaks corresponding to the protons of the vinyl ester function at 4.56, 4.88 and 7.29 ppm, respectively.

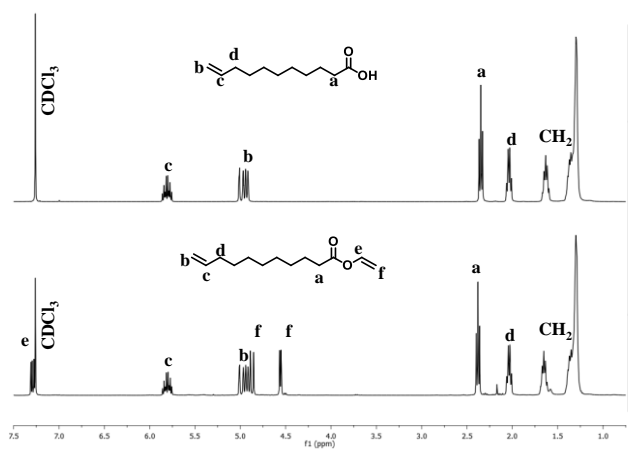

**Fig. 1** $^1$H NMR spectra of undecenoic acid (top) and vinyl undecenoate (bottom).

The transesterification reaction of the trehalose was then performed using *Candida antarctica* Lipase B immobilized on acrylic resin (CALB) also referred as Novozyme 435. This enzyme is the most used in the literature due to its stability in organic solvents and its high efficiency.[12,40] The reaction was conducted in dry methyl-tetrahydrofurane at 45°C.

The formation and purity of trehalose diester was proved by means of NMR techniques (see **Fig.2** and **Fig. S1**) as well as ESI-MS analysis (see **Fig S2**). The $^1$H NMR analysis in DMSO-d$_6$ (**Fig 2**) confirmed that the ester functions were only localized on the primary position. Indeed, peaks corresponding to protons located nearby the primary alcohols were shifted from 3.48 and 3.53 ppm to 4.24 and 4.03 for protons in 6-position and from 3.64 to 3.88 ppm for protons in 5-position, respectively. No shifts of the protons nearby the secondary alcohols were observed.

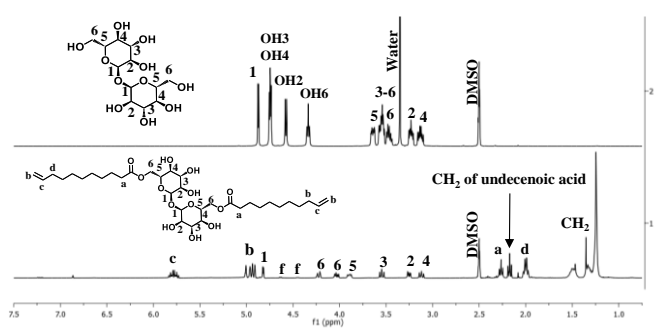

**Fig. 2** $^1$H NMR spectra of trehalose (top) and trehalose diundecenoate before purification (bottom).

**Synthesis of polymers by ADMET polymerization**

ADMET polymerization is performed on α,ω-diene monomers to produce unsaturated polymers. It is a step-growth polymerization driven by the release of ethylene. When the melting points of the monomers are not too high (<100°C), the polymerization can be conducted in bulk and under vacuum to remove the ethylene from the reaction medium. Thus, the equilibrium of the reaction is driven towards the polymer formation. In our case, the melting point of the trehalose diundecenoate is around 145°C, value incompatible with the catalyst stability to perform the polymerization in bulk. ADMET homopolymerization of trehalose diundecenoate and its copolymerization with undecenyl undecenoate (see $^1$NMR spectrum, **Fig S3**), were thus conducted in THF for 24 hours at 45°C (Scheme 2).

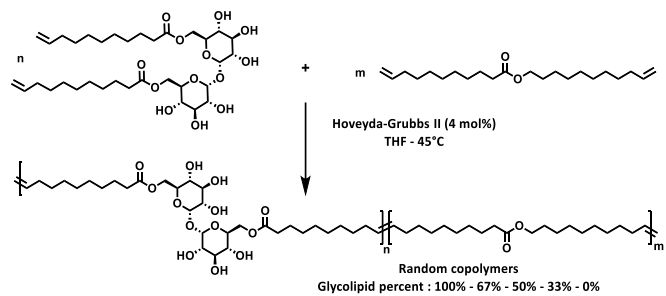

**Scheme 2** Synthesis of polymers by ADMET (co)polymerization of trehalose diundecenoate.

First, the reactivity of trehalose diundecenoate toward ruthenium-based metathesis catalysts was investigated. Indeed, the efficiency of the catalyst depends on several factors, like the solubility of the monomer or the tolerance of the catalyst toward chemical functions present on the monomers. In order to select the best catalyst, trehalose diundecenoate was homopolymerized in THF with 4 mol.% of Grubbs 1$^{st}$ (G1), Grubbs 2$^{nd}$ (G2), Hoveyda-Grubbs 1$^{st}$ (HG1) and Hoveyda-Grubbs 2$^{nd}$ (HG2) generation metathesis catalysts, respectively (**see Fig. S4**). The completion of the reaction was determined by $^1$H NMR; for each polymer the disappearance of the terminal double bond peaks at 5.00 and 5.8 ppm and the appearance of the internal double bond peak at 5.35 ppm confirmed the polymer formation (**see Fig. S5**). After precipitation in cold methanol, the poly(trehalose undeceneoate)s were analyzed by SEC, in DMF; data are given in **Table S6** and **Fig. S6**. It is worth noting that molar masses and dispersity provided by SEC should not be taken as absolute values as the SEC calibration was carried out using polystyrene standards Except for HG1, all catalysts produced poly(trehalose undeceneoate)s with reasonable molar masses and expected dispersities around 2. However, the higher reactivity of HG2 and its tolerance to moisture led us to retain this catalyst for the copolymerization series. In this last case, undecenyl undecenoate was added in different proportions from 33 mol.% to 67 mol.% to prepare random copolymers. The completion of the polymerization reactions and the composition of each copolymer were evaluated by $^1$H NMR. The completion of the reaction was confirmed with the disappearance of the terminal double bond signal (5.00 and 5.8 ppm) and the appearance of the internal double bond signal (5.35 ppm). The copolymer composition was evaluated using the 4.2 ppm chemical shift of protons in 6-position on the trehalose (**Fig. S7**). The experimental polymer composition was in good agreement with the theoretical one (**Fig. S7** and **Table 1**). MALDI-TOF also supported the formation of the copolymers (**Fig. S8 to S11**). The molar masses of the (co)polymers obtained by ADMET polymerization were determined by SEC in DMF. The polymers were found insoluble in THF at room temperature when the proportion of glycolipid in the polymer was higher than 33 mol.%; however the latter were all soluble in DMF. Data are summarized in **Table 1**. The SEC traces, in DMF, of the copolymers obtained by ADMET polymerization are shown in **Fig. 3**.

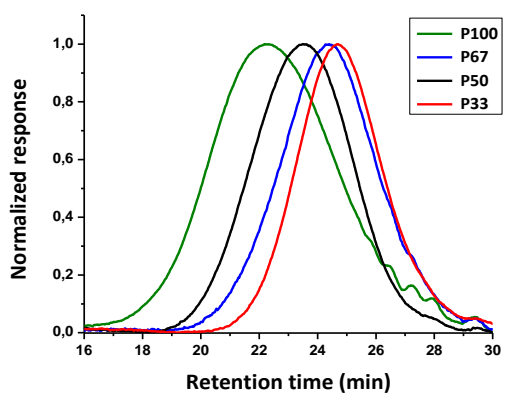 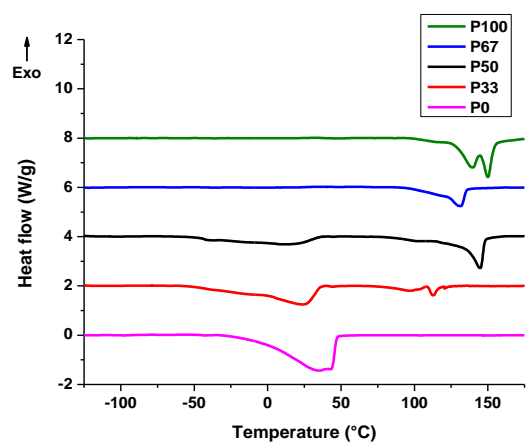

**Fig. 3** SEC traces of the copolymers synthesized by ADMET from trehalose diundecenoate and undecenyl undecenoate with different proportions of trehalose diundecenoate (33–100%). SEC performed in DMF with LiBr Polystyrene calibration.

**Fig. 4** DSC curves of the polymers synthesized by ADMET from trehalose diundecenoate and undecenyl undecenoate with different proportions of trehalose diundecenoate (0–100%).

It can be noticed that the molar masses of the copolymers increase with the glycolipid content. The low molar masses obtained, notably in the case of low content of trehalose diundeconate (P33), is explained by the formation of a fraction of cyclic oligomers during ADMET polymerization, as evidenced by MALDI-TOF analysis (**Fig S11**). However, in the frame of this study, polymers with high content in glycolipid moiety were mostly targeted.

**Thermal properties and bulk phase separation morphology of copolymers**

The thermal stability of the homopolymers and copolymers was evaluated by thermogravimetry (TGA) under nitrogen at a heating rate of 10 °C.min$^{-1}$. Two thermal decompositions can be observed, the first one corresponding to the saccharide segment around 330°C and the second one corresponding to the lipid part around 420°C (**Fig. S12**). The thermal behavior of the (co)polymers was determined by DSC. All analyses were carried out under a nitrogen atmosphere with a 10 °C.min$^{-1}$ heating rate. The DSC traces are shown in **Fig. 4**. All the polymers evidenced a semi-crystalline behavior. No glass transition temperature (Tg) could be detected by DSC even after quenching the melted samples in liquid nitrogen. P0 and P100 present a melting temperature (Tm) of 145 °C for undecenyl undecenoate and 35 °C for trehalose diundecenoate, respectively.

When 33 mol.% of glycolipid is incorporated in the copolymer (P33), two melting temperatures are observed, one corresponding to the "lipidic part" around 25°C and one at 110°C corresponding to the "saccharide part". At a ratio of 50 mol.% of incorporated glycolipid (P50), the melting temperature of the "saccharide part" increases to 145°C and the melting temperature of the fatty counterpart stays around 20°C. Above 50 mol.% (P67), only one melting temperature could be detected at 130°C. The observation of the two melting transitions from the copolymer segments is clear evidence that a phase separation occurred in the bulk material. More insight about the resulting structure was obtained by X-ray scattering.

Small-angle X-ray scattering experiments were thus performed at different temperatures, below and after the previously observed phase transitions, on the (co)polymers to elucidate their solid-state self-assembled structure. **Fig. 5 (a)** shows X-ray spectra of the series of (co)polymers measured at 30 °C. It can be noticed that X-ray spectrum of the "fatty" homopolymer (P0) only shows scattering peaks in the wide-angle region (q = 15.20 and 16.6 nm$^{-1}$) corresponding to characteristic distance around 0.40 nm (d=2π/q), which could be attributed to the lateral distance between aliphatic chains, in agreement with previously reported data.[41] The diffractogram of the homopolymer synthesized from trehalose undecenoate (P100) shows characteristic peaks in the low-angle region, with a wave-vector periodicity that is characteristic to a lamellar packing (q = 2.45, 4.9, 7.35, 9.8 and 12.5 nm$^{-1}$). The first intense low-angle reflection observed at q = 2.45 nm$^{-1}$ is assigned to a long period of 2.56 nm. This long period could be attributed to the distance between two trehalose units in the polymer backbone, in agreement with earlier work[22] on polysophorolipids. A schematic representation of the packing model is represented in **Fig. 6**.

**Table 1** Summary of characteristics of the polymers synthesized by ADMET from trehalose diundecenoate and undecenyl undecenoate with different proportions of trehalose diundecenoate (0–100%).

| Polymers | Trehalose diundecenoate feed (mol.%) | Trehalose diundecenoate incorporated (mol.%)[a] | $\overline{M}_n$[b] (g.mol$^{-1}$) | Đ[b] | $T_m$ (°C) | ΔH (J/g) | $T_{d5\%}$ (°C) |
|---|---|---|---|---|---|---|---|
| P100 | 100 | 100 | 13200 | 2.1 | 156 | 19.57 | 275 |
| P67 | 67 | 61 | 6000 | 1.7 | 130 | 13.60 | 280 |
| P50 | 50 | 45 | 9600 | 1.6 | 12 - 145 | 9.2 – 9.9 | 278 |
| P33 | 33 | 27 | 5500 | 1.5[d] | 24 - 110 | 32.67 – 5.43 | 275 |
| P0 | 0 | 0 | 2500[c] | 1.2[c] | 35 | 83 | 380 |

[a] Determined by $^1$H NMR spectroscopy. [b] Estimated by SEC in DMF with LiBr, PS calibration. [c] Determined by SEC in THF, PS calibration.

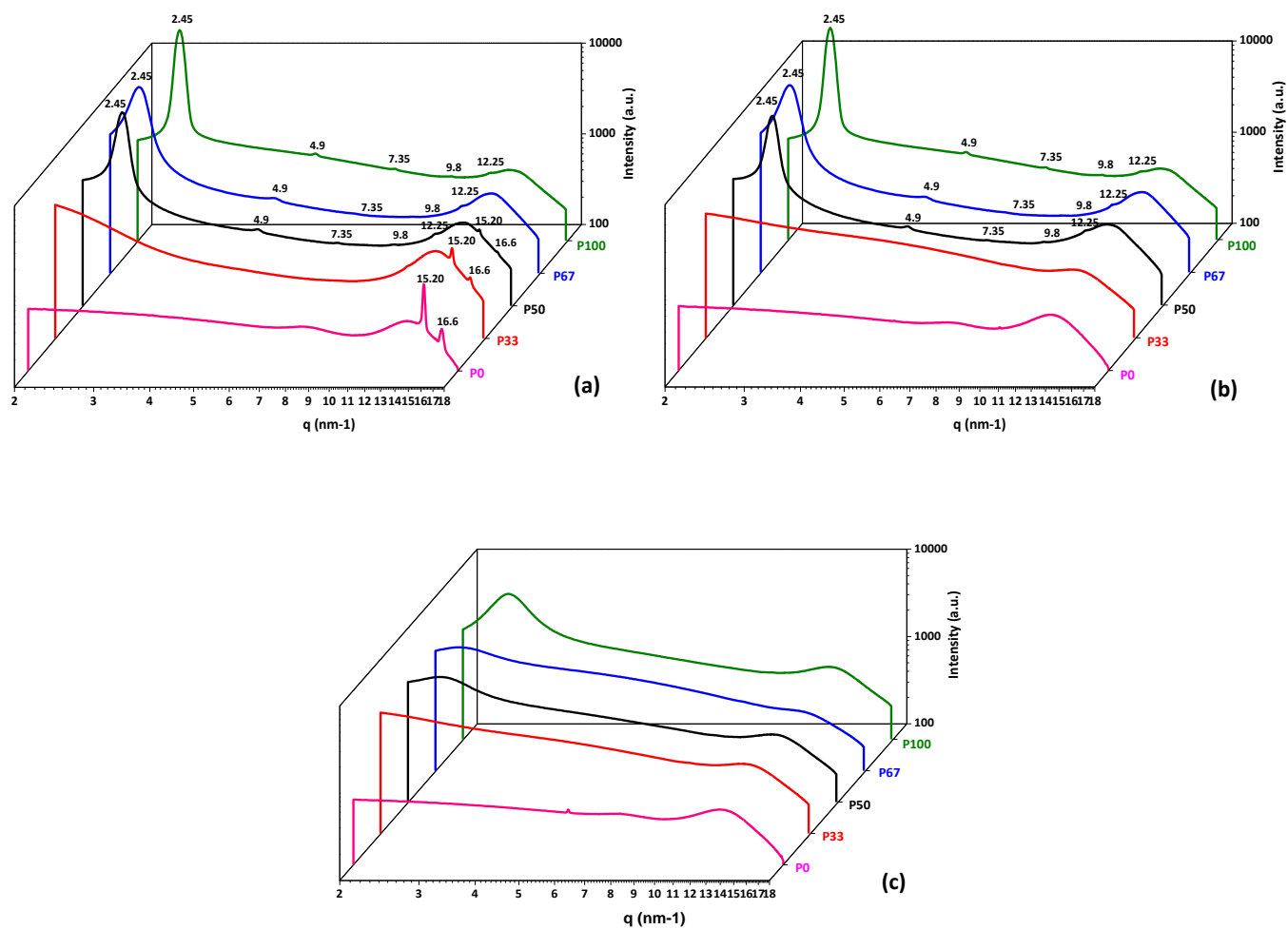

**Fig. 5** X-ray spectra recorded at 30°C (a), 100°C (b) and 160°C (c) of the polymers synthesized by ADMET from trehalose diundecenoate and undecenyl undecenoate with different proportions of trehalose diundecenoate (0–100%).

However, no characteristic peak corresponding to short distance that may reflect the lipid phase crystallization is observed on the diffractogram. This feature could be explained by the presence of the trehalose units in the polymer backbone that can create inter-chain spacing, which may prevent the crystallization of the lipid domains. When 33 % of glycolipid is incorporated in the copolymer (P33), only the crystallization of the lipidic phase at short distance is observed. However, a double melting point was observed on the DSC curves. The too small amount of trehalose units incorporated in the polymer backbone cannot induce a lamellar organization allowing its observation by X-ray scattering.

With 50 % of glycolipid in the copolymer (P50), the two types of organizations are observed. There are enough saccharide and lipid segments to allow the observation of the two crystallizations.

When 67 % of glycolipid is incorporated in the copolymer structure (P67), the lamellar organization of trehalose is only observed. The large amount of saccharide present in the backbone probably prevents the crystallization of the lipidic phase, as it is the case for the homopolymer synthesis with 100% of glycolipid (P100).

**Fig. 5 (b)** and **Fig. 5 (c)** show X-ray spectra of the (co)polymers in bulk respectively at 100 °C and 160 °C. At 100 °C (**Fig. 5 (b)**), the short distance reflections (d = 0.40 nm) disappeared in the cases of homopolymer P0 and of the copolymers P33 and P50 containing 33 % and 50 % of glycolipid respectively, showing the disappearance of the lipidic crystalline domains, in agreement with DSC data.

At 160 °C (**Fig. 5 (c)**), above the melting temperatures of both segments, the long period reflections (d=2.56 nm), corresponding to the lamellar packing between saccharide units along the chains, disappears. That corresponds to the second melting observed by DSC.

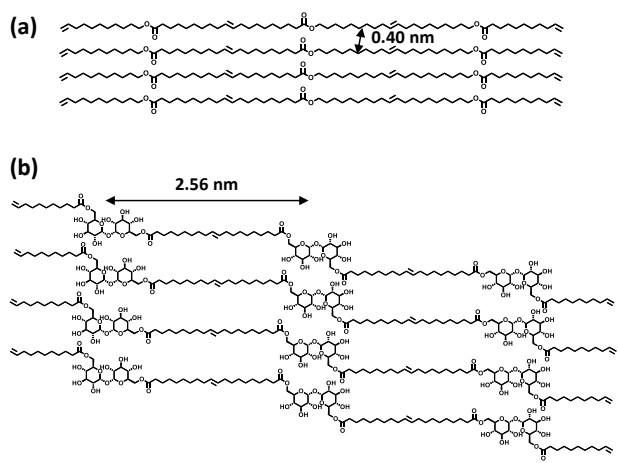

**Fig. 6** Packing model showing lipid crystallization of the homopolymer P0 synthesized with 0 % of glycolipid (a) and lamellar packing of the homopolymer P100 synthesized with 100 % of glycolipid (b).

### Self-assembly in water: nanoparticles formation and characterization

The self-assembly behavior in water of the (co)polymers were studied using a solvent-displacement (or nanoprecipitation) method. To induce self-assembly, solutions of polymers in DMF (a good solvent for both the lipid and saccharide moieties) were dialyzed against water, which is a non-solvent for the lipidic part, thus inducing spontaneously their assembly. Milky solutions were obtained after dialysis. The concentration of the polymer dispersion in water was adjusted to 0.5 mg.mL$^{-1}$ by dilution. The resulting assemblies were characterized by transmission electron microscopy (TEM) and dynamic light scattering (DLS). The hydrodynamic diameters and polydispersity indices (PDIs) of the polymer assemblies were determined and are summarized in

**Table 2**.

As shown in **Fig. 7**, TEM images showed micellar self-assemblies. DLS analysis (see **Fig S13**) showed that all the polymers containing the glycolipid (glycopolyesters) unit in the polymer chains exhibit only one relaxation time in agreement with a monomodal distribution, proving that these unsaturated glycopolyesters are able to nicely self-assemble in water by nanoprecipitation. The homopolymer P0, which was synthesized without glycolipid, precipitated after dialysis and, as expected, no colloïdally stable assembly was observed.

**Table 2** Hydrodynamic diameters and polydispersity indices (PDIs) measured by DLS of the assemblies formed from the polymers synthesized by ADMET from trehalose diundecenoate and undecenyl undecenoate with different proportions of trehalose diundecenoate (0–100%).

| Polymers | $d_H$ (nm) | PDI |
|---|---|---|
| **P100** | 200 | 0.01 |
| **P67** | 160 | 0.22 |
| **P50** | 160 | 0.07 |
| **P33** | 160 | 0.11 |
| **P0** | - | - |

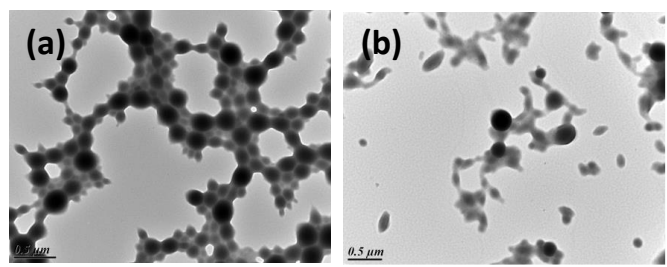

**Fig. 7** TEM images of the (co)polymers (a) P100 and (5) P50 after self-assembly.

DLS measurements showed that the resulting self-assemblies have hydrodynamic diameters ($d_H$) of 200 nm for the homopolymer obtained from the trehalose diundecenoate (P100) and $d_H$ of about 160 nm for all the copolymers (P67, P50 and P33) and with relatively low PDI (below 0.25). The nanoparticles size appears to be mostly independent of the polymer composition and thus dominated by the self-assembly process. In addition, the nanoparticle dispersions appeared to be stable for at least several days.

## Conclusions

A series of random copolymers were synthesized from ADMET polymerization of trehalose diundecenoate and undecenyl undecenoate. These novel glycolipid-polyesters exhibit unique structural organization due to the lipidic and carbohydrate part that induce crystallization and ordered packing within the materials of the saccharide and the lipid moieties. In bulk, the resulting polymers exhibit two distinct melting points around 30°C and 130°C. Moreover, these amphiphilic copolymers could self-assemble in nano-scale particles in water. It was observed that the morphology and the size of the particles were independent on the glyco-to-lipid percentage in the copolymers, thus allowing modulating their average hydrophobicity (hydrophilicity) without changing their size and colloidal stability.

## Acknowledgements

The authors thank University of Bordeaux, Bordeaux INP, CNRS, Aquitaine Regional Council and ITERG for the financial support of this research. They would also thank Centre de Recherche Paul Pascal (CRPP) for X-ray analyses, Centre d'Etude Stucturale et d'Analyse des Molécules Organiques (CESAMO) for MALDI-TOF, Equipex Xyloforest ANR-10-EQPX-16 XYLOFOREST for flash chromatography and Bordeaux Imaging Center (BIC) for TEM.

## Notes and references

# ADMET polymerization of glycolipids: solid-state and self-assembly properties of unsaturated glyco-polyesters

G. Hibert,[a,b] E. Grau,[a,b] D. Pintori,[c] S. Lecommandoux,[a,b] and H. Cramail*[a,b]

[a] University of Bordeaux, Laboratoire de Chimie des Polymères Organiques, UMR 5629, IPB/ENSCBP, 16 avenue Pey-Berland, F-33607 Pessac Cedex, France.
[b] Centre National de la Recherche Scientifique, Laboratoire de Chimie des Polymères Organiques UMR 5629 F-33607 Pessac Cedex, France.
[c] ITERG, 11 rue Gaspard Monge, F-33600 Pessac, France.

## Experimental and Supporting Information

## Experimental Methods

### Synthesis

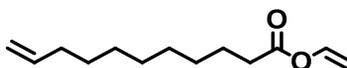

***Vinyl undecenoate (1) (transvinylation of undecenoic acid):*** Undecenoic acid (1 eq) and a 15 eq. excess of vinyl acetate (VAc) was poured in a vial for the microwave reactor. Then, the palladium acetate (0.05 eq.), and the potassium hydroxide (0.10 eq.) were added and the resulting reaction mixture was stirred under microwave at 60 °C for 2 h. The reaction mixture was diluted in DCM and then filtrated over celite to remove the palladium acetate, before removing the solvent with a rotary evaporator. The resulting residue was purified by silica gel flash chromatography using an elution gradient of 2-5% MeOH in DCM to give the vinyl undecenoate. Yield: 95 %. $^1$H NMR (DMSO-d$_6$, 400MHz, δ (ppm)): 7.29 (m, 1H, =C$\underline{H}$-OCO-), 5.81 (m, 1H, -C$\underline{H}$=CH$_2$), 4.97 (m, 2H, C$\underline{H}_2$=CH-), 4.88 (d, 1H, CH$_2$=C$\underline{H}$-OCO-), 4.56 (d, 1H, CH$_2$=C$\underline{H}$-OCO-

), 2.37 (t, 4H, -CH₂-COO-), 2.04 (m, 4H, -CH₂-CH=CH-), 1.67 (m, 4H, -CH₂-CH₂-COO-), 1.30 (m, 20H, aliphatic -CH₂-).

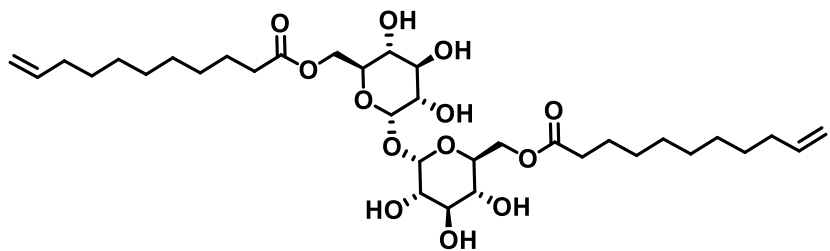

***Trehalose diundecenoate (2) (enzymatic esterification of trehalose):*** The lipase (2.8 g) was added to a mixture of trehalose (3 g, 9 mmol), vinyl ester (6.8 g, 22 mmol, 2.5 eq) in dry acetone (40 mL). The reaction mixture was stirred at 45 °C for 72 hr. After the reaction time, THF was added to well dissolve the diesters of trehalose, then the reaction mixture was filtered and solvent was removed in rotary evaporator. The obtained crude product was purified by silica gel flash chromatography using an elution gradient of 5-25% methanol in EtOAc-DCM (1:1) to give pure trehalose diesters as white solids. Yield: 50 %. ¹H NMR (DMSO-d₆, 400MHz, δ (ppm)): 5.78 (m, 2H, -CH=CH2), 5.04 (d, 2H, -OH, H4), 4.94 (m, 4H, CH2=CH-), 4.89 (d, 2H, -OH, H3), 4.82 (d, 2H, -CH-, H1), 4.76 (d, 2H, -OH, H2), 4.21 (d, 2H, -CH-, H6), 4.04 (m, 2H, -CH-, H6), 3.89 (m, 2H, -CH-, H5), 3.55 (m, 2H, -CH-, H3), 3.26 (m, 2H, -CH-, H2), 3.13 (m, 2H, -CH-, H4), 2.27 (t, 4H, -CH2-COO-), 2.01 (m, 4H, -CH2-CH=CH-), 1.51 (m, 4H, -CH2-CH2-COO-), 1.33-1.25 (m, 20H, aliphatic -CH2-) ).

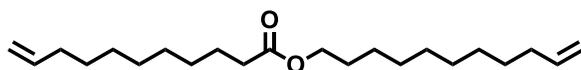

***Undecenyl undecenoate (3) (transesterification):*** Undecenol (12.8 g, 0.08 mol.) was blended with 10-methylundecenoate (15 g, 0.08 mol.). TBD (5% mol.) was added as a catalyst. The reaction was performed under a nitrogen flow at 120°C for 2h, then the temperature was increased to 160 °C for 2h more under dynamic vacuum. Purification over silica gel flash chromatography was performed using cyclohexane/ethyl acetate 94/6 eluent. Yield: 76%. ¹H NMR (400MHz, CDCl₃, δ (ppm)): 5.8 (m, 1H, -CH=CH₂), 4.9 (m, 2H, CH₂=CH-), 4.0 (t, 2H, -CH₂COO-), 2.2 (t, 2H, -COOCH₂-), 2.0 (m, 4H, -CH₂-CH=CH-), 1.5-1.2 (m, 26H, aliphatic -CH₂-).

**Synthesis of polymers**

*ADMET polymerization*

***Homopolymers (P0 and P100):*** Into a flame-dried Schlenk flask equipped with bubbler, undecenyl undecenoate (P0) (0.1 g, 0.3 mmol.) or trehalose diundecenoate (P100) (0.2 g, 0.3 mmol.) dried over-night under vacuum was dissolved in 2 mL of dry THF. Hoveyda-Grubbs 2nd generation metathesis catalyst (4 mol.%) was added and the reaction mixture was stirred under nitrogen atmosphere for 24 h at 45 °C. Then, 3 ml of ethyl vinyl ether were introduced into the flask to quench the reaction. The final glyco-polyester or  was purified by precipitation in cold methanol.

***Copolymers (P33, P50 and P67):*** Into a flame-dried round Schlenk flask, trehalose diundecenoate (0.2 g, 0.3 mmol.) and the corresponding amount of undecenyl undecenoate (see **Erreur ! Source du renvoi introuvable.**) were mixed and dried over-night under vacuum. Then, the diene monomers were dissolved in 2 mL of dry THF and Hoveyda-Grubbs 2nd generation (4 mol.%) was added. The reaction mixture was stirred at 45 °C under nitrogen atmosphere for 24 h. Then, 3 mL of ethyl vinyl ether and 4 mL of THF were introduced to the flask. The final copolymers were purified by precipitation in cold methanol.

**Preparation of polymer self-assemblies in water**

The nano-particles were prepared using solvent displacement method by dialysis. (Co)polymers were solubilized in DMF and polymer solutions were then poured into dialysis membranes with molecular weight cutoff of 1kD. The membranes were beforehand soaked for 15 min and rinsed with deionized water. The membranes were then submerged in 2 L of deionized water and dialyzed under gentle magnetic stirring for 12 h. After 2 h and 4 h, the dialysis solution was replaced by fresh deionized water. At the end of the dialysis, milky solutions present inside the dialysis membranes were recover.

## Supporting Information

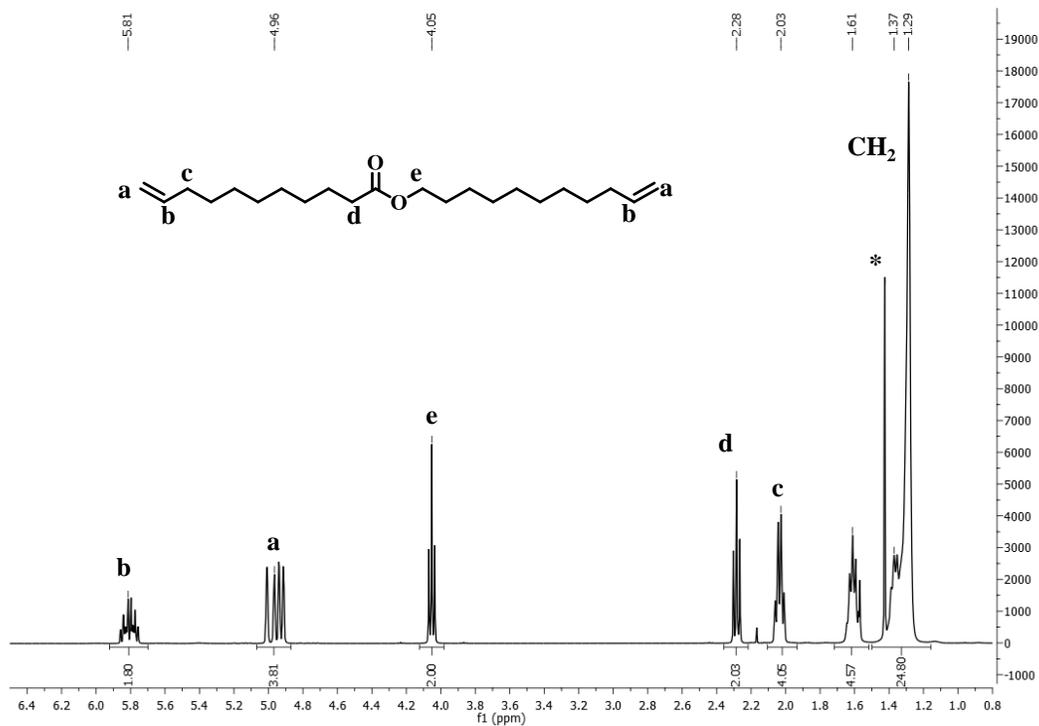

**Fig. S1** ¹H NMR spectrum of undecenyl undecenoate performed in CDCL₃.

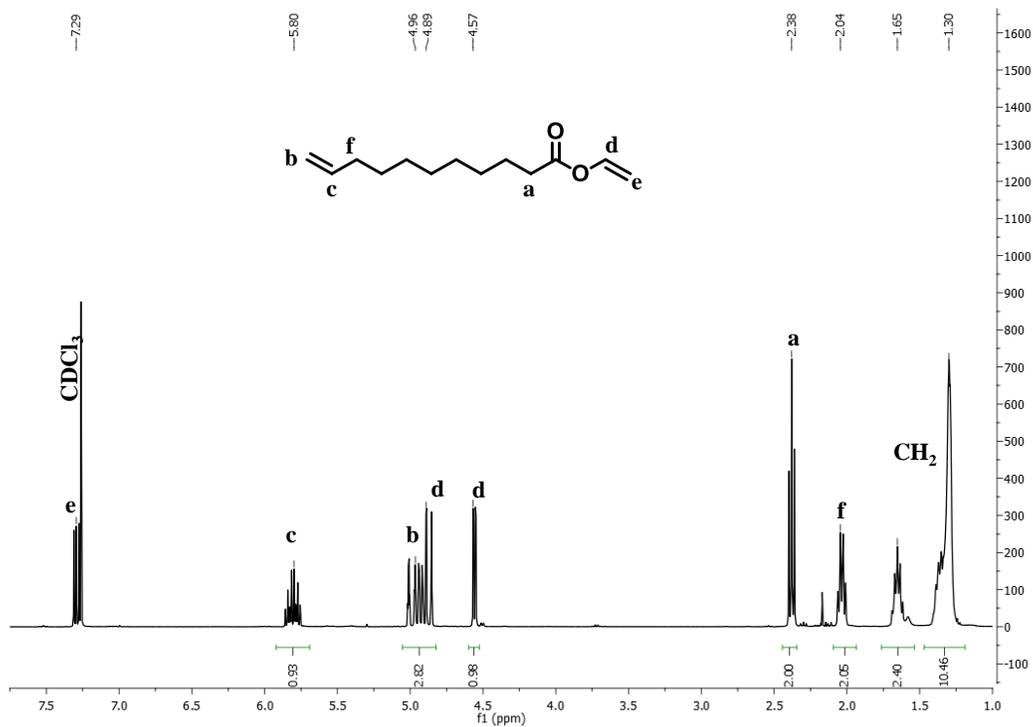

**Fig. S2** ¹H NMR spectrum of vinyl undecenoate performed in CDCL₃.

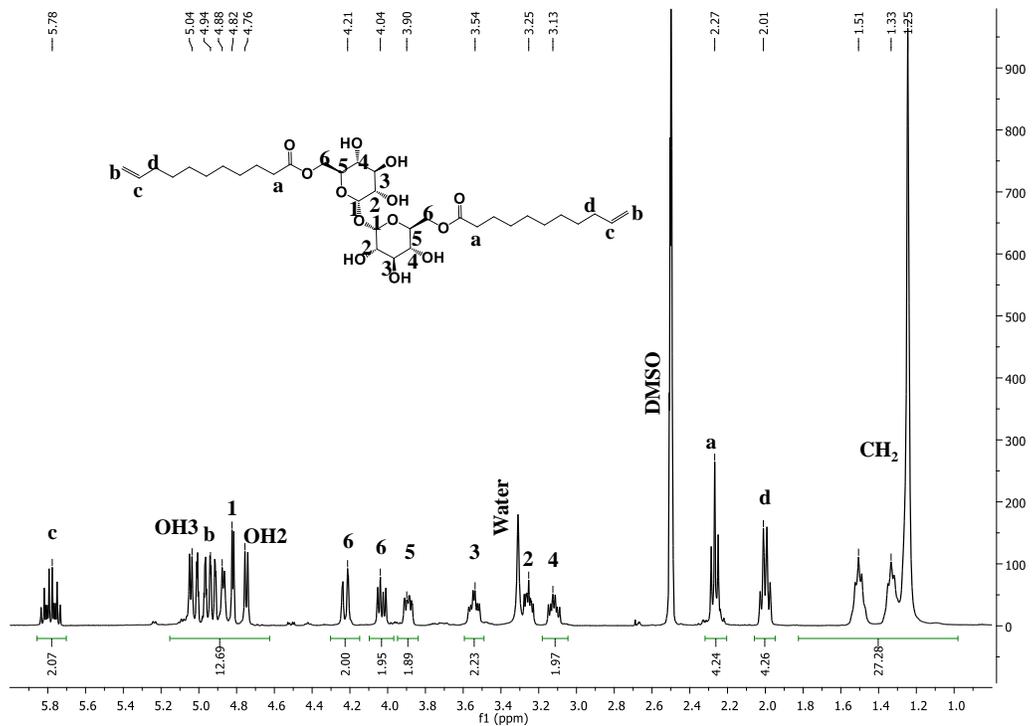

**Fig. S3** ¹H NMR spectrum of pure trehalose diundecenoate performed in DMSO-d₆.

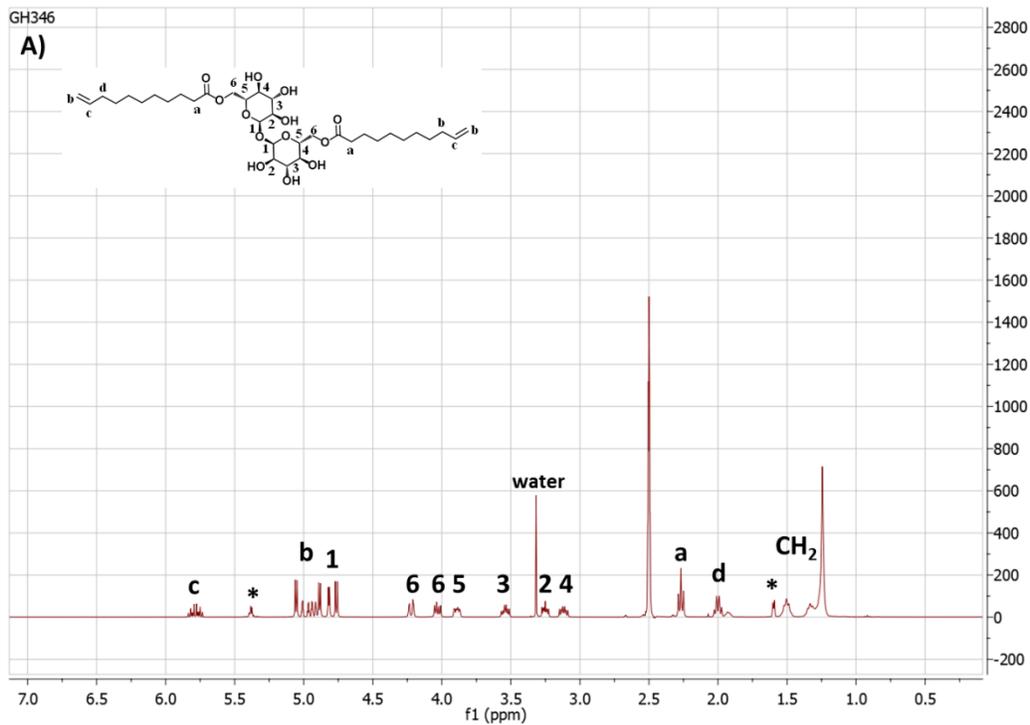

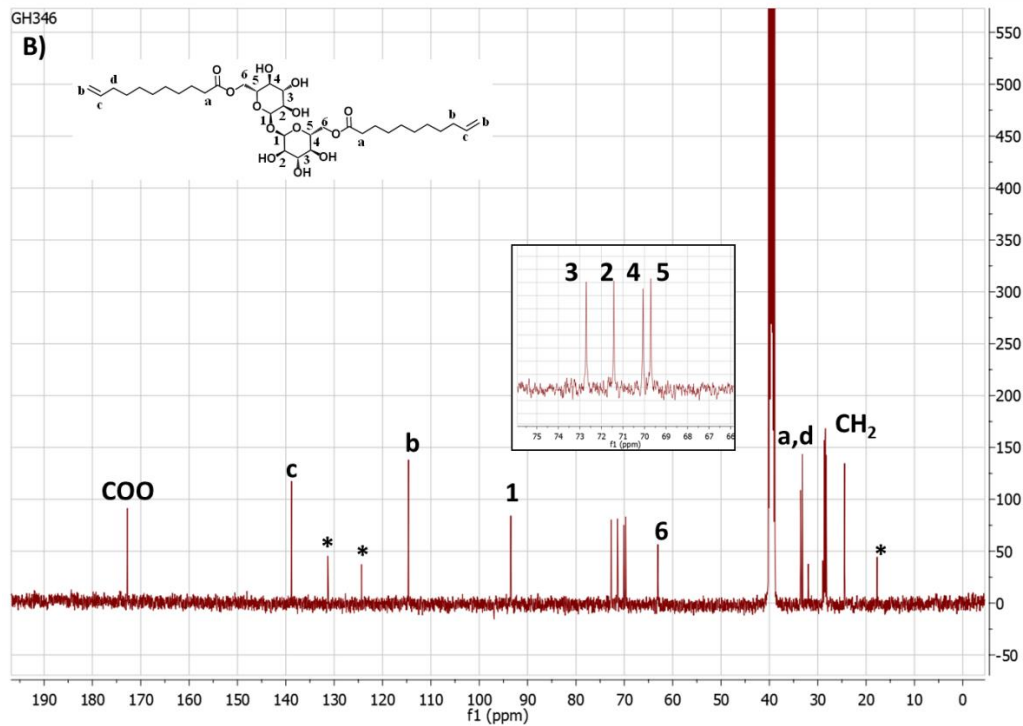
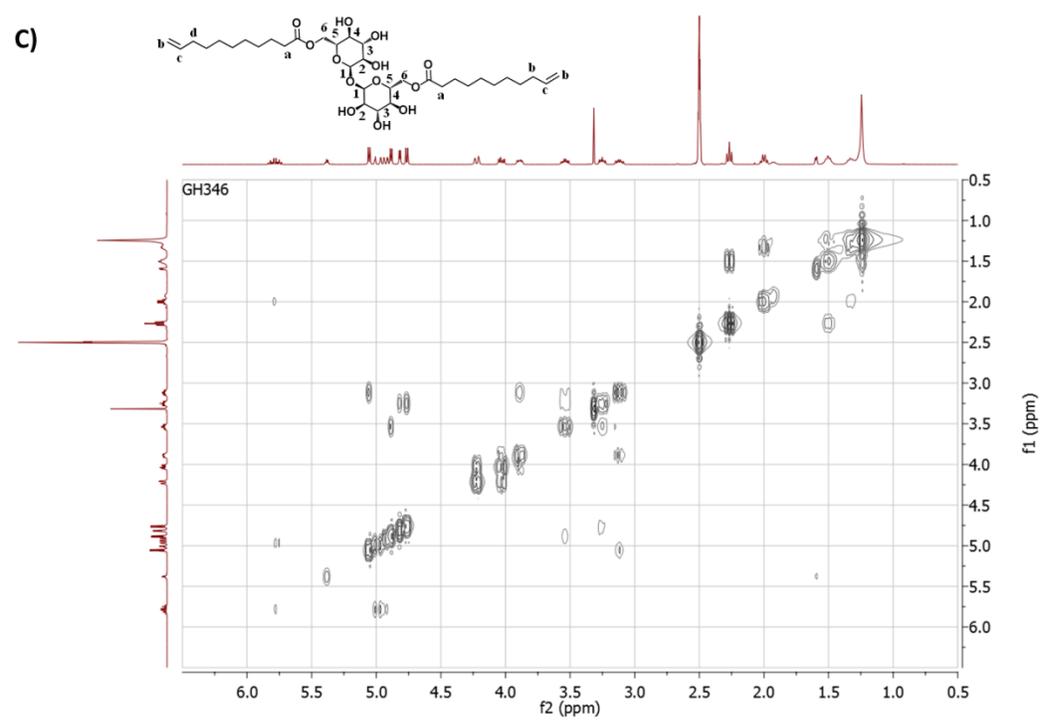

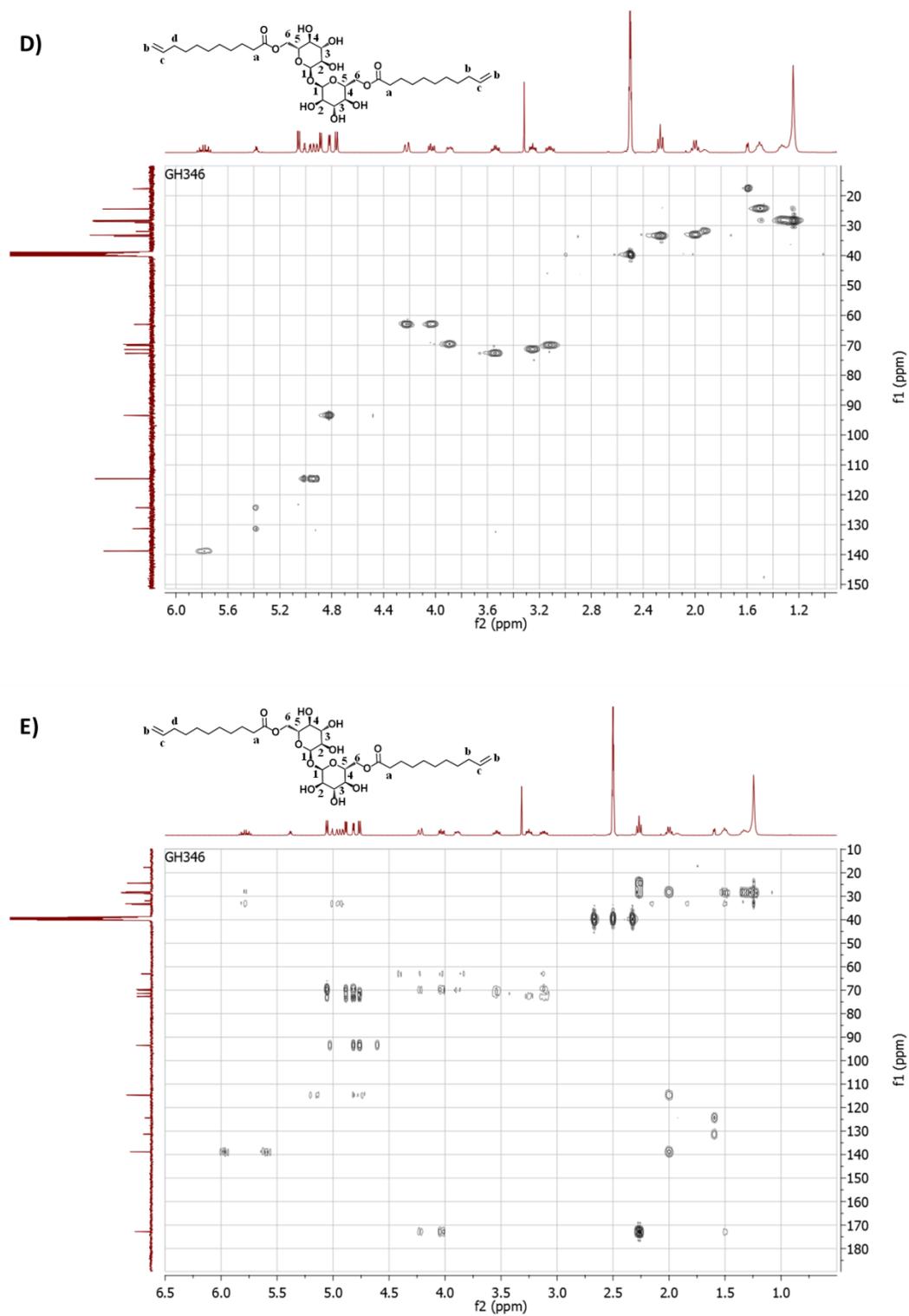

**Fig. S4** NMR spectra of trehalose diundecenoate performed in DMSO-$d_6$. A) $^1$H NMR, B) $^{13}$C NMR, C) $^1$H-$^1$H COSY NMR, D) $^1$H-$^{13}$C HSQC NMR, E) $^1$H-$^{13}$C HMBC NMR. * corresponded to isomerize double bonds of trehalode diundecenoate

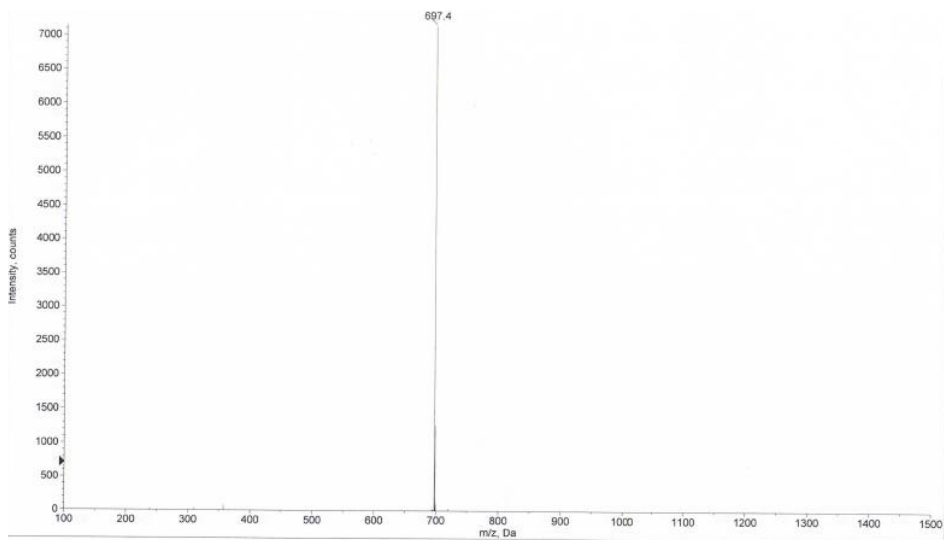

Fig. S5 ESI-MS of trehalose diundecenoate

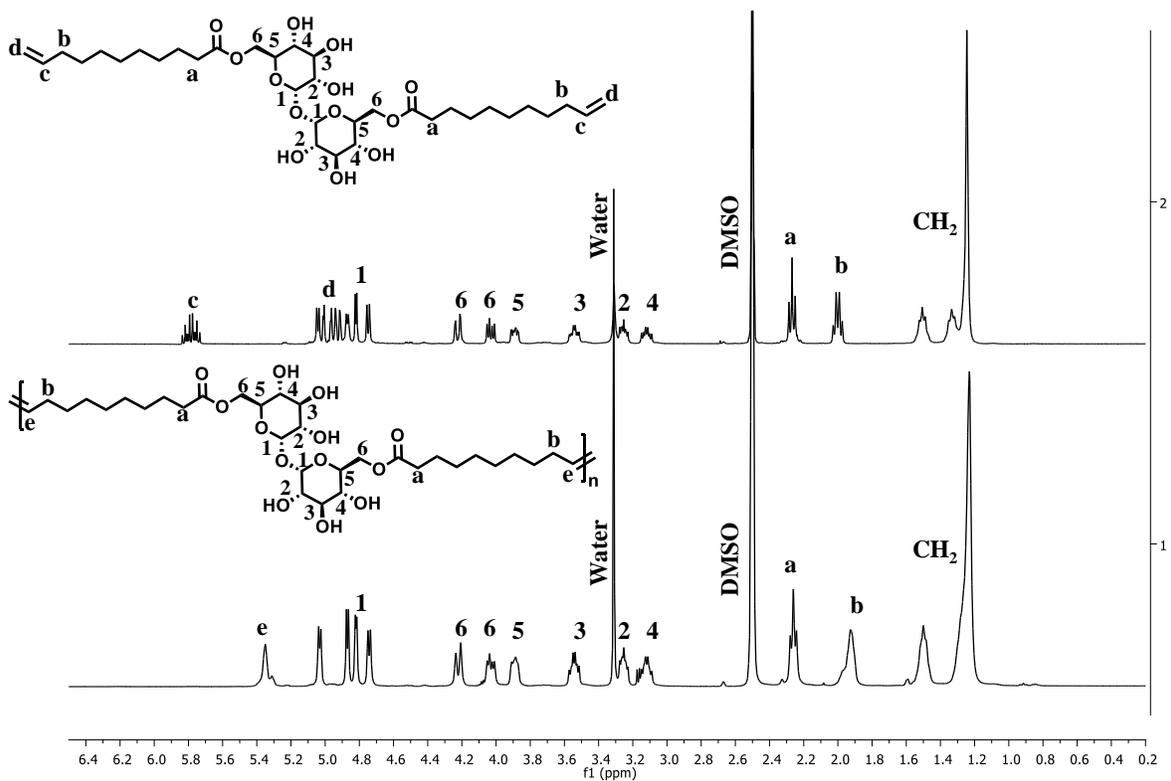

**Fig. S6** Stacked [1]H NMR spectra of trehalose diundecenoate (top) and the homopolymer P100 synthesized by ADMET from trehalose diundecenoate (bottom) performed in DMSO-$d_6$.

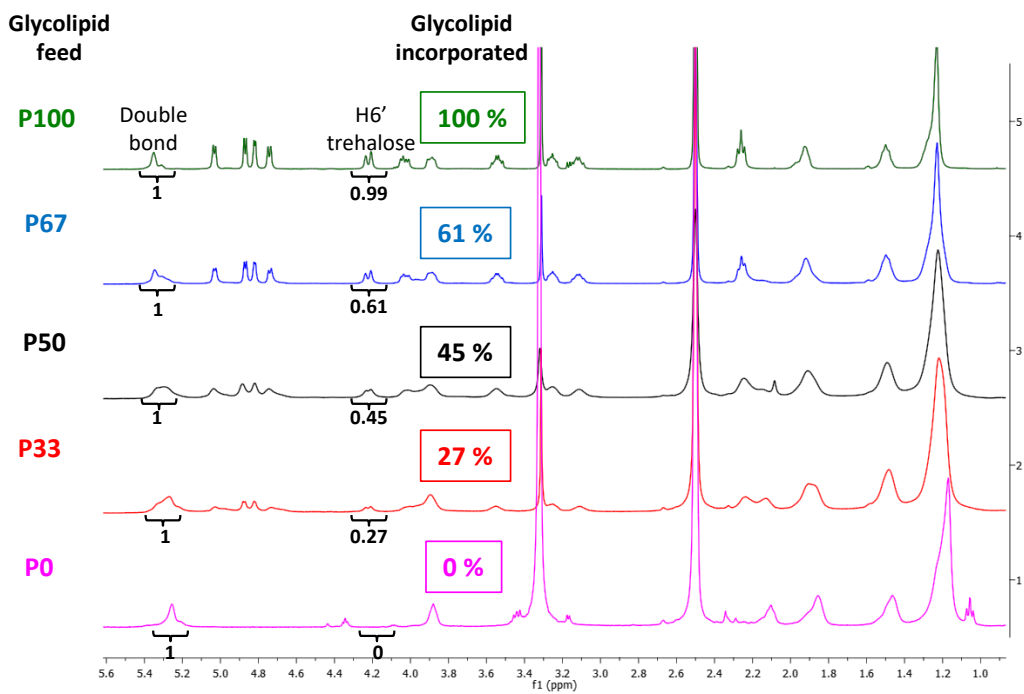

**Fig. S7** $^1$H NMR spectra in DMSO of the copolymers obtained by ADMET, with ratio of trehalose undecenoate incorporated (100 % = 100 % trehalose diundecenoate).

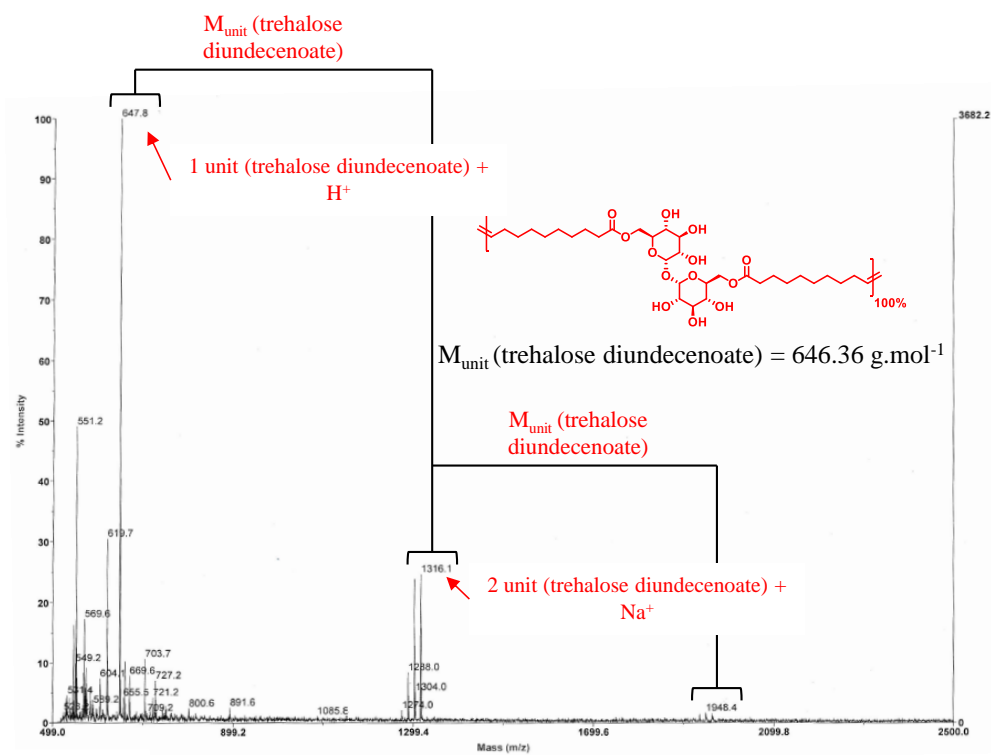

**Fig. S8** MALDI-TOF analysis of P100 (Matrix 2,5-dihydroxybenzoic acid (DHB))

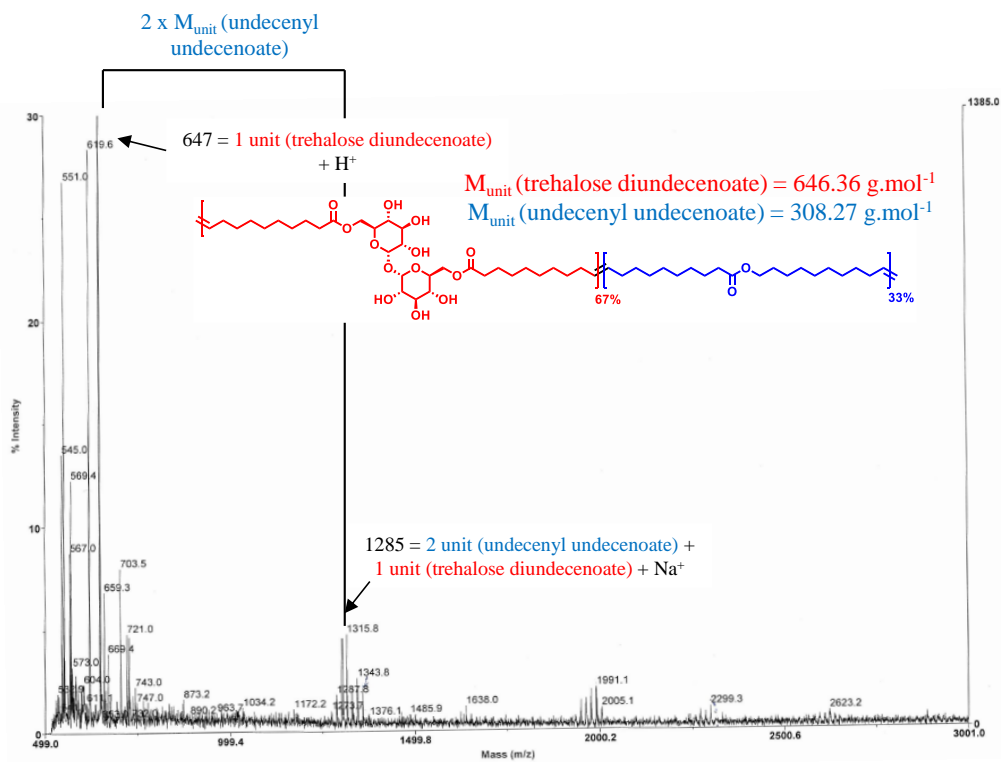

**Fig. S 9** MALDI-TOF analysis of P67 (Matrix 2,5-dihydroxybenzoic acid (DHB))

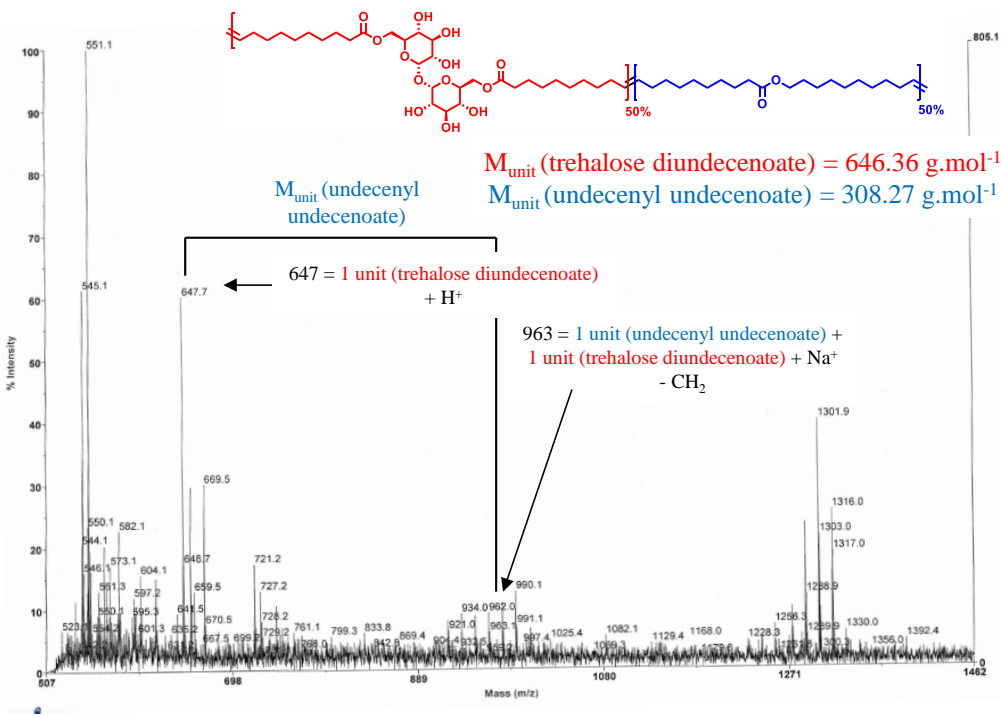

**Fig. S10** MALDI-TOF analysis of P50 (Matrix 2,5-dihydroxybenzoic acid (DHB))

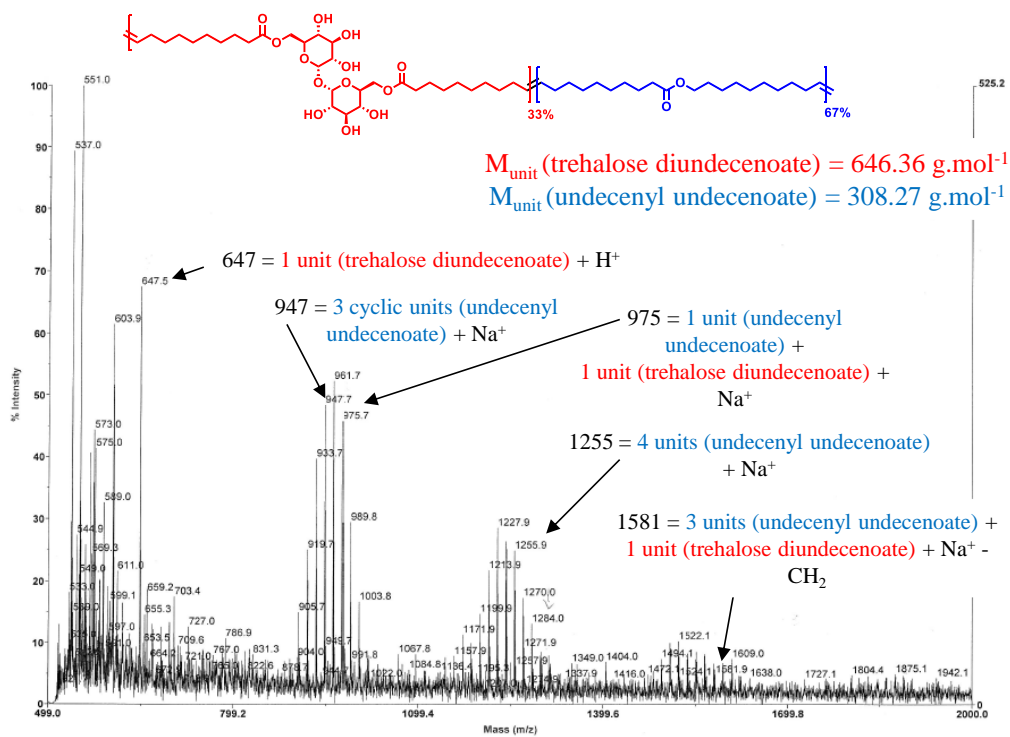

**Fig. S11** MALDI-TOF analysis of P33 (Matrix 2,5-dihydroxybenzoic acid (DHB))

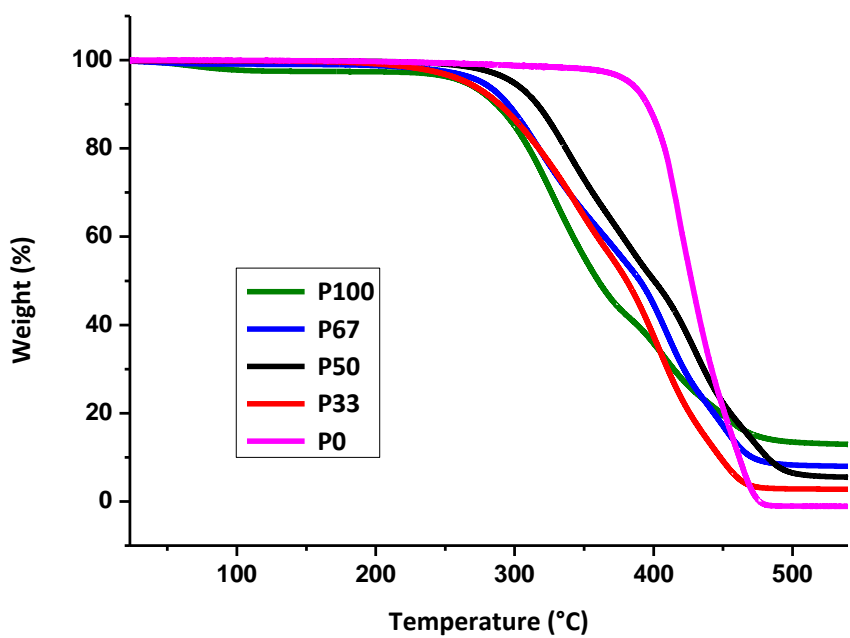

**Fig. S12** TGA curves of (co)polymers obtained by ADMET polymerization.

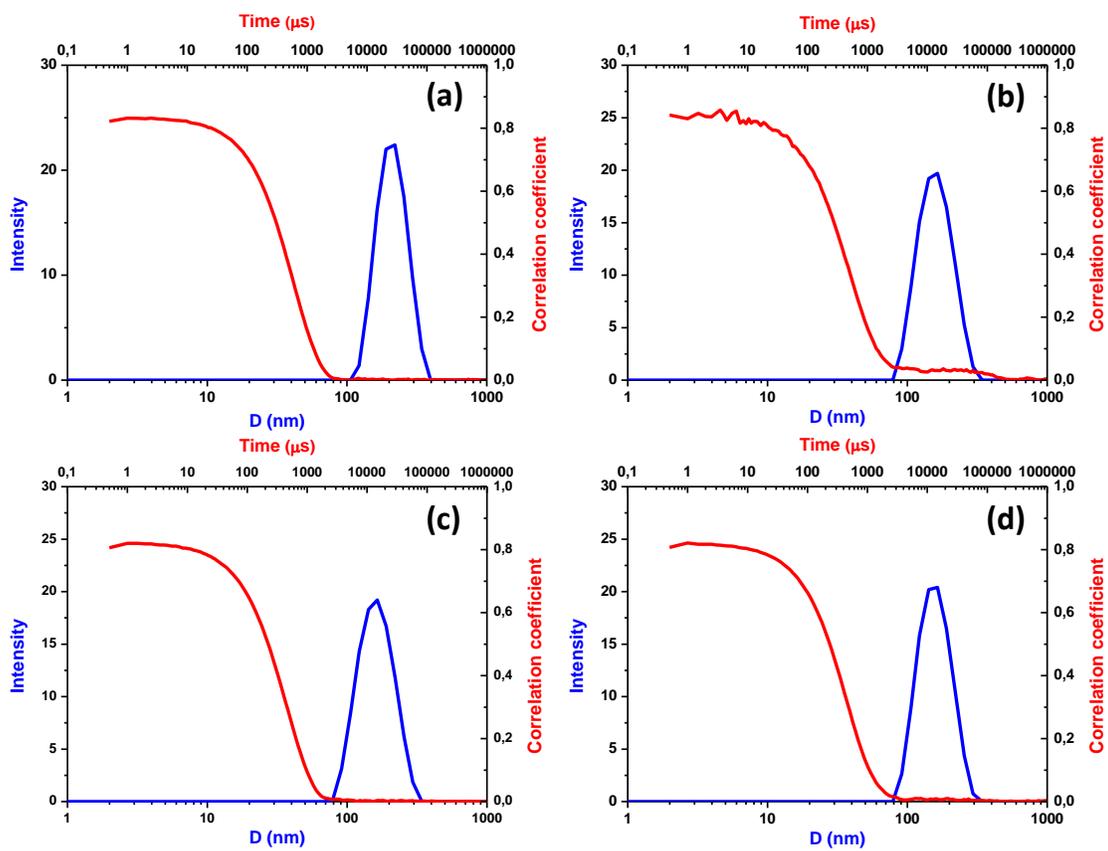

**Fig. S13** Auto-correlation functions (red) and relaxation time distribution (blue) (determined at 90°) scattering obtained from DLS for assemblies in water: (a) P100, (b) P67, (c) P50, (d) P33